\documentclass[aps,12pt,superscriptaddress,tightenlines,preprint,floatfix]{revtex4-1}
\usepackage{amsmath,amssymb,graphicx,bm}
\newcommand{\nn}{\nonumber} 

\newcommand{\R}{\mathbb{R}}
\providecommand{\abs}[1]{\left\lvert#1\right\rvert}
\providecommand{\norm}[1]{\lVert#1\rVert}
\DeclareMathOperator{\li2}{\textrm{Li}_2}
\allowdisplaybreaks

\begin{document}
\title{An iterative method for spherical bounces}
\author{Roman V. Buniy}
\email{roman.buniy@gmail.com}
\affiliation{Schmid College of Science and Technology, Chapman University, Orange, CA 92866}
\date{\today}
\begin{abstract}
We develop a new iterative method for finding approximate solutions for spherical bounces associated with the decay of the false vacuum in scalar field theories.
The method works for any generic potential in any number of dimensions, contains Coleman's thin-wall approximation as its first iteration, and greatly improves its accuracy by including higher order terms. 
\end{abstract}
\maketitle

\section{Introduction} \label{section_introduction}

Although a scalar field theory is often used only as a first approximation towards a more precise description of a physical system, it can nevertheless reveal significant properties of the system. 
Among many features of scalar field theories that have been extensively studied, those describing special kinds of solutions --- solitons, instantons and bounces --- are particularly interesting.
Bounces, for example, are related to stability properties of classical and quantum configurations in such theories.
This relation exists because a stable classical state may be only metastable quantum-mechanically.
The instability is realized by allowing a metastable state to tunnel to a stable state via a quantum barrier penetration or to thermally climb over the barrier in order to arrive at the stable state.
These processes are widely used in describing various physical systems ranging from phase transitions in solids to bubble formation in cosmological inflaton fields~\cite{Voloshin,Frampton:1976kf,Frampton:1976pb}.

Due to its prevalence in theoretical models, the tunneling in scalar field theories should be thoroughly understood and accurate approximation methods for its equations should be developed.
Toward the latter goal, we specifically focus on approximation methods for the bounce in Euclidean field theories.
The bounce itself and the first approximation for it (called the thin-wall approximation) was introduced by Coleman~\cite{Coleman:1977py}.
Callan and Coleman~\cite{Callan:1977pt} developed the first quantum corrections for the bounce.
Coleman, Glaser and Martin~\cite{Coleman:1977th} proved that, for a wide class of potentials, spherically-symmetric solutions to equations of motion are the solutions with the lowest action.
Coleman and De Luccia~\cite{Coleman:1980aw} considered modification to the thin-wall approximation due to gravitational effects and showed that increasing the effects of gravity can render the false vacuum stable.
Since these seminal works on the bounce, significant progress has been made in improving accuracy and generality of the thin-wall approximation, for example, for a restricted class of polynomial potentials~\cite{Shen:si} and non-polynomial potentials~\cite{Samuel:mz}.
Quantum corrections for the bounce were also enhanced and the decay rate of the false vacuum was obtained, for example, in the one loop effective action calculations~\cite{Baacke:2003uw}, \cite{Dunne:2005rt}.    
Further studies showed intricate properties of  gravitational bounces, \cite{Samuel:1991dy}, \cite{Masoumi:2016pqb}.

Here we propose a new method of approximate solutions for the multidimensional spherical bounce.
The method starts with the thin-wall approximation as its first step and proceeds to higher orders iteratively with fast convergence and high accuracy.
Analysis of the problem from a new perspective demonstrates some universal properties of the bounce.
The method is not restricted to only certain types of potentials or dimensions of space, and we demonstrate its computational power with the general fourth-order polynomial potential.
We find that the approximation works well beyond its intended range of applicability. 

\section{Spherical bounces} \label{section_bounces}

Consider a scalar field theory defined by the action
\begin{align}
  S=\int_{\R^{n+1}}dx\biggl[\frac{1}{2}\norm{\nabla\phi}^2 +U(\phi)\biggr],
  \label{action}
\end{align}
where $\phi\colon\R^{n+1}\to\R$ is a scalar field, $U\colon\R\to\R$ is a potential function (which we assume to be continuously differentiable), $\nabla$ is the gradient operator in $\R^{n+1}$, and $\norm{\cdot}$ is the norm in $\R^{n+1}$.
The corresponding Euler-Lagrange equation is
\begin{align}
  \nabla^2\phi=\frac{dU}{d\phi},\label{equation_of_motion}
\end{align}
where $\nabla^2$ is the Laplace operator in $\R^{n+1}$.

Let $\phi=\phi_1$ be a minimum of $U$.
For the theory defined by the action $S$, the solution $\phi=\phi_1$ is classically stable, but its quantum stability depends on the type of the minimum.
If the minimum is absolute, the solution is stable and is called a true vacuum; if the minimum is relative, the solution is unstable and is called a false vacuum.
At zero temperature, which we assume throughout our analysis, a false vacuum decays into a true vacuum by the process of barrier tunneling.
To study the simplest example of such tunneling, we choose $U$ with two minima and one maximum, set the absolute minimum at $\phi=\phi_{-}$, the relative minimum at $\phi=\phi_{+}$ and the relative maximum at $\phi=\phi_{*}$.

For computational convenience, and without any loss of generality, certain conditions can be imposed on the function $U$.
To derive them, we start with the analog of \eqref{equation_of_motion} for the variables $(y,\psi,V)$, 
\begin{align}
  &\nabla_y^2\psi(y)=\frac{dV(\psi(y))}{d\psi(y)},\label{equation_of_motion_new_coordinates}
\end{align}
and transform these to the variables $(x,\phi,U)$ according to
\begin{align}
  &y=ax,\\
  &\psi(y)=b+c\phi(x),\\
  &V(\psi)=g+hU(\phi),
  \label{coordinate_transformation}
\end{align}
where $b$ and $g$ are constants, and $a$, $c$ and $h$ are nonzero constants.
As a result, \eqref{equation_of_motion_new_coordinates} becomes
\begin{align}
  &\nabla_x^2\phi(x)=\frac{a^2 h}{c^2}\frac{dU(\phi(x))}{d\phi(x)},\label{equation_of_motion_transformed}
\end{align}
which coincides with \eqref{equation_of_motion} if the constraint $a^2 h=c^2$ holds.
As we have freedom to choose five coefficients $a$, $b$, $c$, $g$ and $h$ subject to this constraint, this is equivalent to having freedom to impose four independent conditions on the function $U$.
For the first two conditions, we set $\phi_{-}$ and $\phi_{+}$ to take particular values, and it is convenient to choose $\phi_{-}=-\phi_{+}$ and $\phi_{+}>0$.
For the third condition, we set $U(\phi_{+})=0$.
Finally, for the fourth condition, we set 
\begin{align}
  \min_\phi{(U(\phi)+U(-\phi))}=U(\phi_{+})+U(\phi_{-}), \label{u_bound}
\end{align}
the reason for the form of which will become clear in Sec.~\ref{subsection_orders}.
We denote $U(\phi_{-})=-\epsilon$ for some $\epsilon>0$ and write \eqref{u_bound} as
\begin{align}
  \min_\phi{(U(\phi)+U(-\phi))}=-\epsilon. \label{u_bound_epsilon}
\end{align}
If the condition \eqref{u_bound_epsilon} is satisfied without the fourth restriction on the function $U$ (as is the case for the general fourth-order polynomial potential considered in Sec.~\ref{section_polynomial}), we can impose one additional condition on $U$.

\begin{figure}[h]
  \begin{center}
    \includegraphics[width=468pt]{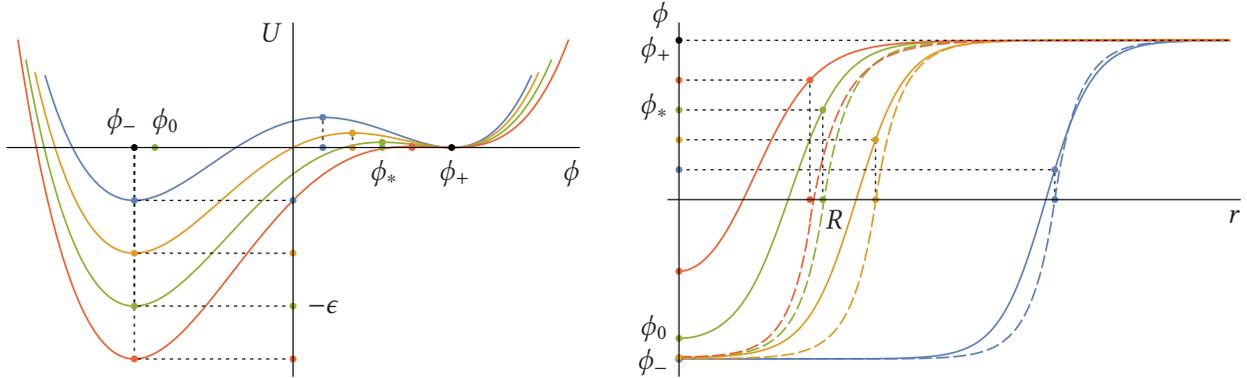}
  \end{center}
  \caption{Left: examples of the potential function $U$ of different degrees of asymmetry with two minima and one maximum.
    To simplify comparison of the resulting solutions, we choose, without any loss of generality (see Sec.~\ref{section_bounces}), the absolute minimum at $\phi=\phi_{-}$, the relative minimum at $\phi=\phi_{+}$ and $U(\phi_{+})=0$ for all potentials.
    We set the relative maximum at $\phi=\phi_{*}$ and the turning point at $\phi=\phi_0$, where $\phi_{-}\le\phi_{0}\le\phi_{*}$; for $n=0$, we have $\phi_{-}=\phi_0$.
    The energy-density difference between the true and false vacua  $\epsilon=U(\phi_{+})-U(\phi_{-})$ characterizes the degree of asymmetry of $U$.
  The quantities $\phi_0$, $\phi_*$ and $-\epsilon$ are labeled only for one curve.
  The potential functions shown here are precisely those for which the corresponding solutions and their approximations are shown in the right part of this figure and in Figs.~\ref{figure_u_tilde_0} and \ref{figure_phi_0}. 
  Right: the spherical half-bounce solutions for $n=3$ and the potentials in the left part of this figure.
The solid curves are for the exact numerical solutions and the dashed curves are for the approximate analytic solutions given by \eqref{phi_small_r}, \eqref{phi_large_r}, \eqref{phi_0_constant} and \eqref{B_constant}.
The quantities $\phi_0$, $\phi_*$ and $R$ are labeled only for one curve, which corresponds to the curve labeled in the left part of this figure.
  Each color of curves in this figure and in Figs.~\ref{figure_u_tilde_0} and \ref{figure_phi_0} represents the same value of $\epsilon$. 
  }
  \label{figure_u_phi}
\end{figure}

See Fig.~\ref{figure_u_phi} for examples of $U$.

A solution $\phi$ of \eqref{equation_of_motion} is called a bounce if it satisfies the boundary conditions  
\begin{align}
  \lim_{x_0\to\pm\infty}\phi(x)&=\phi_{+},\label{bondary_condition_1}\\
  \lim_{x_0\to 0}\frac{\partial\phi(x)}{\partial x_0}&=0,\label{bondary_condition_2}
\end{align}
where $x=(x_0,x_1,\dotsc,x_n)$.
These conditions mean that the field starts at the false vacuum $\phi=\phi_{+}$ at $x_0=-\infty$, reaches the turning point $\phi=\phi_0$ at $x_0=0$, bounces back and finally reaches the false vacuum at $x_0=\infty$.
We are interested in a bounce for which the action is finite.
In addition to $U(\phi_{+})=0$ that we have set above, finiteness of the action also requires
\begin{align}
  \lim_{\sum_{i=1}^n x_i^2\to\infty}\phi(x)=\phi_{+}.
  \label{bondary_condition_3}
\end{align}

For a large class of potentials $U$, spherically symmetric solutions of \eqref{equation_of_motion} are the solutions with the lowest action~\cite{Coleman:1977th} and we will be concerned here only with these.
A function $\phi\colon[0,\infty)\to\R$ of the Euclidean distance $r=(\sum_{i=0}^n x_i^2)^{1/2}$ satisfies the equation
\begin{align}
  \frac{d^2\phi}{dr^2}+\frac{n}{r}\frac{d\phi}{dr}=\frac{dU}{d\phi}
  \label{phi_equation}
\end{align}
and the boundary conditions
\begin{align}
  \lim_{r\to\infty}\phi(r)&=\phi_{+},\label{boundary_condition_r_1}\\
  \lim_{r\to 0}\frac{d\phi(r)}{dr}&=0.\label{boundary_condition_r_2}
\end{align}
The solution of \eqref{phi_equation}, \eqref{boundary_condition_r_1} and \eqref{boundary_condition_r_2} is a spherical half-bounce since the field starts at the turning point $\phi=\phi_0$ at $r=0$ and reaches the false vacuum $\phi=\phi_{+}$ at $r=\infty$.  
Upon carrying out the angular integration, the action \eqref{action} becomes
\begin{align}
  S=A\int_0^\infty d r\,r^n\biggl[\frac{1}{2}\Bigl(\frac{d\phi}{dr}\Bigr)^2 +U(\phi)\biggr], \ A=\frac{2\pi^{(n+1)/2}}{\Gamma((n+1)/2)}, \label{action_r}
\end{align}
where $A$ is the area of an $(n+1)$-dimensional unit sphere.

The classical analog of \eqref{equation_of_motion} is a particle moving in the potential $-U$ and subject to a viscous damping force; see e.g.~\cite{Coleman:1977py}.
The viscous damping always dissipates energy and so for the bounce the field at the turning point $\phi=\phi_0$ still has lower potential energy than the final field $\phi_{+}$, i.e., $U(\phi_0)<U(\phi_{+})$, but also $\phi_0>\phi_{-}$ and $U(\phi_0)>U(\phi_{-})$; see Fig.~\ref{figure_u_phi}.

\section{The exact solution for $\boldsymbol{\mathrm{n}=0}$} \label{section_exact}

In this section we derive the exact solution of the field equation \eqref{phi_equation} for $n=0$,
\begin{align}
  \frac{d^2\phi}{dr^2}=\frac{dU}{d\phi};
  \label{phi_equation_0}
\end{align}
although elementary, it serves as a starting point for our approximation solution of \eqref{phi_equation} for $n\ge 1$ for which no exact solution is known.

We first note that according to \eqref{phi_equation_0}, the particle moves without viscous damping, so that its energy is conserved.
Multiplying both sides of \eqref{phi_equation_0} by $d\phi/dr$, integrating over $r$ and using the boundary condition \eqref{boundary_condition_r_2} together with $U(\phi_{-})=-\epsilon$, we find
\begin{align}
  \frac{1}{2}\Bigl(\frac{d\phi}{dr}\Bigr)^2=U(\phi)+\epsilon.
  \label{}
\end{align}
This implies
\begin{align}
  \frac{dr}{d\phi}=(2(U(\phi)+\epsilon))^{-1/2},
  \label{bounce_0_dr_dphi}
\end{align}
where we have chosen the positive sign of the square root for the positive half-bounce for which $\phi(0)<\phi(\infty)$. 
The integration of \eqref{bounce_0_dr_dphi} with the boundary condition $r(\phi_{-})=0$ (because $\phi_{-}=\phi_0$ for $n=0$) now gives
\begin{align}
  r(\phi)=\int_{\phi_{-}}^\phi d\phi_1\,(2(U(\phi_1)+\epsilon))^{-1/2}.
  \label{bounce_0}
\end{align}
Since $r(\phi)$ is a monotonically increasing function, we can define the radius $R=r(\phi_{*})$ (so that $\phi(R)=\phi_{*}$) and write \eqref{bounce_0} in the form which is more convenient for subsequent calculations,
\begin{align}
  &r(\phi)=R+\int_{\phi_{*}}^\phi d\phi_1\,(2(U(\phi_1)+\epsilon))^{-1/2}, \label{bounce_0_with_R}\\
  &R=\int_{\phi_{-}}^{\phi_{*}} d\phi_1\,(2(U(\phi_1)+\epsilon))^{-1/2}.  \label{bounce_0_R}
\end{align}
Equation \eqref{bounce_0_with_R} is the one-dimensional instanton centered at $r=R$, for which the action \eqref{action_r} becomes
\begin{align}
  S=2\int_0^\infty dr\,(2U(\phi(r))+\epsilon)=2\int_{\phi_{-}}^{\phi_{+}}d\phi\frac{2U(\phi)+\epsilon}{(2(U(\phi)+\epsilon))^{1/2}},
  \label{}
\end{align}
where we used \eqref{bounce_0_dr_dphi} twice.

\section{The thin-wall approximation} \label{section_thin_wall_approximation}

\subsection{A power series expansion} \label{subsection_power_series_approximation}

While there is no exact solution of the field equation \eqref{phi_equation} for $n>0$, one approximate solution is well-known \cite{Coleman:1977py}.
This so-called thin-wall approximation applies when the potential function $U$ is nearly symmetric and consequently the energy-density difference between the false and true vacua $\epsilon=U(\phi_{+})-U(\phi_{-})$ is small.
The small asymmetry of the function $U$ implies that viscous damping for the particle from the mechanical analogy is small and that the turning point for the bounce is near the absolute minimum, $\phi_0\approx\phi_{-}$.
It turns out that $\phi_0-\phi_{-}$ is exponentially small in $\epsilon$.

The particle moving in such a potential spends a long time in the neighborhood of $\phi=\phi_0$ before it crosses the potential valley.
The crossing happens somewhere between $\phi=\phi_0$ and $\phi=\phi_{+}$ and it is convenient to take $\phi=\phi_*$ as the center of the valley; see the left part of Fig.~\ref{figure_u_phi}.
It is also said that the wall separating the regions of the false and true vacua is located at $r=R$; see the right part of Fig.~\ref{figure_u_phi}.  
Since it takes a long time for the particle to reach the wall, it follows that $R$ is large for small $\epsilon$; indeed, we will find that $R=O(\epsilon^{-1})$.
(Note that for $\phi(0)>\phi_0$ the particle does not have enough energy to overcome the friction and to reach the local minimum at $\phi=\phi_{+}$; it oscillates around $\phi=\phi_{*}$ and finally stops there.
For $\phi(0)<\phi_0$, the particle arrives to $\phi=\phi_{+}$ with positive energy at a finite $r$, at which point it starts accelerating towards $\phi\to\infty$.) 
Finally, after crossing the potential valley, the particle spends a long time approaching $\phi=\phi_{+}$ as $r\to\infty$.
To prove the above qualitative statements, in the rest of this subsection we solve \eqref{phi_equation} separately for small $r$ and large $r$ and match the two solutions at $r=R$. 

For small $r$ we have $\phi\approx\phi_{-}$ and so we expand
\begin{align}
  U(\phi)\approx U(\phi_{-})+\frac{1}{2}(\phi-\phi_{-})^2 U''(\phi_{-}), \ \phi\approx\phi_{-}.
  \label{u_small_r}
\end{align}
For this potential the solution of \eqref{phi_equation} satisfying the boundary condition $\phi(0)=\phi_0$ is
\begin{align}
  \phi(r)&\approx\phi_{-}+(\phi_0-\phi_{-})\Gamma(\nu+1)(\tfrac{1}{2}k_{-}r)^{-\nu}I_\nu(k_{-}r)\label{phi_small_r}\\
  &\approx\phi_0+\frac{\phi_0-\phi_{-}}{2(n+1)}(k_{-}r)^2+O(r^3), \ r\to 0,
  \label{phi_small_r_approximation}
\end{align}
where
\begin{align}
  \nu&=\tfrac{1}{2}(n-1),\\
  k_\pm&=(U''(\phi_\pm))^{1/2}
  \label{}
\end{align}
and $I_\nu$ is the modified Bessel function of the first kind.
Since $\phi_0-\phi_{-}$  is exponentially small in $\epsilon$ (which we prove below), the function \eqref{phi_small_r} changes very slowly near $r=0$.
In other words, the particle spends a long time in the neighborhood of $\phi=\phi_0$ before it crosses the potential valley.

For large $r$ we have $\phi\approx\phi_{+}$ and so we expand
\begin{align}
  U(\phi)\approx U(\phi_{+})+\frac{1}{2}(\phi-\phi_{+})^2 U''(\phi_{+}), \ \phi\approx\phi_{+}.
  \label{u_large_r}
\end{align}
For this potential the solution of \eqref{phi_equation} satisfying the boundary condition $\phi(\infty)=\phi_{+}$ is
\begin{align}
  \phi(r)&\approx\phi_{+}-B(k_{+}r)^{-\nu}K_\nu(k_{+}r)\label{phi_large_r}\\
  &\approx\phi_{+}-B(\pi/2)^{1/2}(k_{+}r)^{-n/2}\exp{(-k_{+}r)}\bigl(1+O(r^{-1})\bigr), \ r\to\infty,
  \label{}
\end{align}
where $B>0$ is a constant and $K_\nu$ is the modified Bessel function of the second kind.
As a result, $\phi$ approaches its asymptotic value $\phi_{+}$ exponentially slowly.  

Matching the solutions \eqref{phi_small_r} and \eqref{phi_large_r} and their derivatives at $r=R$, we find
\begin{align}
  \phi_0&\approx\phi_{-}+\frac{(\phi_{+}-\phi_{-})k_{+}(\tfrac{1}{2}k_{-}R)^\nu K_{\nu+1}(k_{+}R)}{\Gamma(\nu+1)\bigl[k_{-}I_{\nu+1}(k_{-}R)K_\nu(k_{+}R)+k_{+}I_{\nu}(k_{-}R)K_{\nu+1}(k_{+}R)\bigr]},\label{phi_0_constant}\\
  B&\approx\frac{(\phi_{+}-\phi_{-})k_{-}(k_{+}R)^\nu I_{\nu+1}(k_{-}R)}{k_{-}I_{\nu+1}(k_{-}R)K_\nu(k_{+}R)+k_{+}I_{\nu}(k_{-}R)K_{\nu+1}(k_{+}R)}.\label{B_constant}
\end{align}
Since $R$ is large for small $\epsilon$ (which we prove in Secs.~\ref{subsection_thin_wall_approximation} and \ref{section_method}), we need the expansions for large $k_{+}R$ and $k_{-}R$,
\begin{align}
  \phi_0 &\approx\phi_{-}+\frac{(\phi_{+}-\phi_{-})k_{+}(2\pi k_{-}R)^{1/2}(\tfrac{1}{2}k_{-}R)^\nu\exp{(-k_{-}R)}}{\Gamma(\nu+1)(k_{+}+k_{-})},\label{phi_0_large_R}\\
  B &\approx\frac{(\phi_{+}-\phi_{-})k_{-}(k_{+}R)^\nu(2k_{+}R/\pi)^{1/2}\exp{(k_{+}R)}}{k_{+}+k_{-}}.
  \label{B_large_R}
\end{align}
As stated above, the difference $\phi_0-\phi_{-}$ is indeed exponentially small for large $k_{-}R$.

To estimate the time it takes for the particle to cross the potential valley, we compute
\begin{align}
  \phi'(R)\approx\frac{k_{+}k_{-}(\phi_{+}-\phi_{-})}{k_{+}+k_{-}}
  \label{}
\end{align}
and see that the transition mostly occurs over the interval
\begin{align}
  \abs{r-R}\le\frac{k_{+}+k_{-}}{2k_{+}k_{-}},
  \label{}
\end{align}
which is much smaller than the length of the interval, $O(R)$, over which the particle moves between the two vacua.
In other words, relative to the whole transition, the passage through the potential valley is very fast.

The approximate solution given by \eqref{phi_small_r}, \eqref{phi_large_r}, \eqref{phi_0_constant} and \eqref{B_constant} is a very poor approximation except for very small $\epsilon$; see the right part of Fig.~\ref{figure_u_phi} for examples.
One direct method to improve it is to include in \eqref{u_small_r} and \eqref{u_large_r} terms of higher order in $\phi-\phi_{-}$ and $\phi-\phi_{+}$, respectively, but, unfortunately, no exact solutions of \eqref{phi_equation} are known for such potentials.
It appears that an alternative method of treating higher order terms in $U$ as small perturbations leads to an approximation which is less accurate and more complicated than the solution we derive in the following sections.
One reason for this is that we can view the above approximations as local (due to their reliance on series expansions around either $\phi=\phi_{-}$ or $\phi=\phi_{+}$), while the thin-wall approximation and our generalization of it have features of global solutions which do not give preference to any particular value of $\phi$.

\subsection{The thin-wall approximation} \label{subsection_thin_wall_approximation}

To continue with the thin-wall approximation, we now choose radii $r_1$ and $r_2$ which are close to $R$ and satisfy $0<r_1<R<r_2<\infty$, and proceed with solving \eqref{phi_equation} in three separate regions: $0\le r\le r_1$, $r_1\le r\le r_2$ and $r_2\le r<\infty$.

For $0\le r\le r_1$ we have $\phi(r)\approx\phi_0\approx\phi_{-}$.

For $r_1\le r\le r_2$ we ignore the friction term $(n/r)(d\phi/dr)$ in \eqref{phi_equation} since viscous damping is small.
Furthermore, since asymmetry of $U(\phi)$ is small, we replace $U(\phi)$ with its even part $U_+(\phi)=\frac{1}{2}(U(\phi)+U(-\phi))$ and find
\begin{align}
  \frac{d^2\phi}{dr^2}\approx\frac{dU_{+}(\phi)}{d\phi}.
  \label{phi_equation_approximation}
\end{align}
According to \eqref{phi_equation_approximation}, the particle's motion is approximately the motion in the potential $-U_{+}(\phi)$ without viscous damping, so that its energy is approximately conserved.
We can now proceed as in Sec.~\ref{section_exact} with only small changes due to different boundary conditions.
Since $U_{+}(\phi_{+})=-\epsilon/2$, instead of \eqref{bounce_0_dr_dphi} we have  
\begin{align}
  \frac{dr}{d\phi}\approx(2U_{+}(\phi)+\epsilon)^{-1/2}.
  \label{bounce_approximation_dr_dphi}
\end{align}
Integrating \eqref{bounce_approximation_dr_dphi} with the boundary condition $r(\phi_0)=0$, we find 
\begin{align}
  &r(\phi)\approx R+\int_{\phi_{*}}^\phi d\phi_1\,(2U_{+}(\phi_1)+\epsilon)^{-1/2}, \label{bounce_approximation}\\
  &R\approx\int_{\phi_0}^{\phi_{*}} d\phi_1\,(2U_{+}(\phi_1)+\epsilon)^{-1/2}. \label{bounce_approximation_R}
\end{align}

For $r_2\le r<\infty$ we have $\phi(r)\approx\phi_{+}$.

To compute the action, we consider contributions to the integral in \eqref{action_r} from the three regions used above,
\begin{align}
  S=A\biggl(\int_0^{r_1}+\int_{r_1}^{r_2}+\int_{r_2}^\infty\biggr) d r\,r^n\biggl[\frac{1}{2}\Bigl(\frac{d\phi}{dr}\Bigr)^2 +U(\phi(r))\biggr].
  \label{action_approximation_sum}
\end{align}
For $0\le r\le r_1$ we have $(d\phi/dr)\approx 0$ and $U(\phi)\approx-\epsilon$ and find that this region contributes approximately
\begin{align}
 -A(n+1)^{-1}r_1^{n+1}\epsilon\approx -A(n+1)^{-1}R^{n+1}\epsilon  
  \label{}
\end{align}
to \eqref{action_approximation_sum}, where we set $r_1\approx R$.

For $r_1\le r\le r_2$ we use $(d\phi/dr)^2\approx 2U_{+}+\epsilon$ and $U\approx U_{+}+\epsilon/2$ and find that this region contributes approximately \begin{align}
  A\int_{r_1}^{r_2} dr\, r^n(2U_{+}+\epsilon)
  \label{}
\end{align}
to \eqref{action_approximation_sum}.
Now setting $r\approx R$ in the integrand, changing to the integration over $\phi$, using $(d\phi/dr)\approx(2U_{+}+\epsilon)^{1/2}$ together with $\phi(r_1)\approx\phi_{-}$ and $\phi(r_2)\approx\phi_{+}$, we find that the region $r_1\le r\le r_2$ contributes approximately
\begin{align}
 AR^n\int_{\phi_{-}}^{\phi_{+}} d\phi\, (2U_{+}+\epsilon)^{1/2}  
  \label{}
\end{align}
to \eqref{action_approximation_sum}.

The contribution of the region $r_2\le r<\infty$ to the action can be ignored since $(d\phi/dr)\approx 0$ and $U(\phi)\approx 0$ there.

Combining the above results, we obtain
\begin{align}
  S\approx -A(n+1)^{-1}R^{n+1}\epsilon+AR^n\int_{\phi_{-}}^{\phi_{+}} d\phi\, (2U_{+}(\phi)+\epsilon)^{1/2}.
  \label{action_approximation_0}
\end{align}
The wall location $r=R$ can be determined from \eqref{bounce_approximation_R} if we know $\phi_0$.
Alternatively, we can find it by maximizing $S$ in \eqref{action_approximation_0} with respect to $R$.
Solving $\partial S/\partial R=0$ for $R$, we find 
\begin{align}
  R\approx n\epsilon^{-1}\int_{\phi_-}^{\phi_+} d\phi\,(2U_{+}(\phi)+\epsilon)^{1/2},
  \label{R_approximation}
\end{align}
substitution of which into \eqref{action_approximation_0} finally gives
\begin{align}
  S\approx A(n+1)^{-1}n^n \epsilon^{-n}\biggl[\int_{\phi_-}^{\phi_+} d\phi\,(2U_{+}(\phi)+\epsilon)^{1/2}\biggr]^{n+1}.
  \label{action_approximation}
\end{align}
Equations \eqref{bounce_approximation} and \eqref{R_approximation} represent the thin-wall approximation for the bounce \cite{Coleman:1977py}.

When compared with standard perturbation methods for differential equations, the thin-wall approximation is rather irregular in its derivation.
Despite the presence of a small parameter in the problem, it is not immediately clear how to proceed with the derivation of higher order corrections.
In the following section we develop a systematic approximation scheme which includes the thin-wall approximation as its first iteration. 

\section{The iterative method}\label{section_method}

\subsection{The effective potential}\label{}

As we saw in Secs.~\ref{section_exact} and \ref{section_thin_wall_approximation}, the thin-wall approximation for the solution of \eqref{phi_equation} for $n>0$ can be obtained from the exact solution of \eqref{phi_equation} for $n=0$ with minimal changes.  
To develop an iterative method for generating approximate solutions of \eqref{phi_equation} for $n>0$, we effectively reduce the problem to the case $n=0$ by rewriting \eqref{phi_equation} in the form
\begin{align}
  \frac{d^2\phi}{dr^2}=\frac{d\tilde{U}}{d\phi}.
  \label{phi_equation_tilde}
\end{align}
Since $\phi$ in the spherical half-bounce solution is restricted to the interval $[\phi_0,\phi_{+}]$, the effective potential $\tilde{U}$ introduced via \eqref{phi_equation_tilde} is defined on the same interval. 
Adding an arbitrary constant to $\tilde{U}$ does not change \eqref{phi_equation_tilde}, and we conveniently choose the constant such that $\tilde{U}(\phi_+)=0$ similarly to $U(\phi_+)=0$ that we set in Sec.~\ref{section_bounces}.
Since the particle moves in the potential $-\tilde{U}$ without viscous damping, its energy is conserved; this implies $\tilde{U}(\phi_0)=0$ since $\tilde{U}(\phi_{+})=0$.

Equations \eqref{phi_equation} and \eqref{phi_equation_tilde} lead to
\begin{align}
  \frac{dU}{d\phi}=\frac{d\tilde{U}}{d\phi}+\frac{n}{r}\frac{d\phi}{dr}.
  \label{u_phi}
\end{align}
Since $dU/d\phi$ and $d\phi/dr$ approach zero and $r$ approaches infinity when $\phi$ goes to $\phi_{+}$, \eqref{u_phi} implies that $\lim_{\phi\to\phi_{+}}(d\tilde{U}/d\phi)=0$.
It also follows that $dU/d\phi>d\tilde{U}/d\phi$ for $\phi<\phi_{+}$ since we consider only positive half-bounces for which $d\phi/dr>0$.
From $U(\phi_{+})=\tilde{U}(\phi_+)$ we now conclude that $U(\phi)<\tilde{U}(\phi)$ for $\phi<\phi_+$.

Since the exact solution of \eqref{phi_equation_0} is available, the similarity between \eqref{phi_equation_0} and \eqref{phi_equation_tilde} leads directly to the iterative solution of \eqref{phi_equation_tilde}.
We proceed as in Sec.~\ref{section_exact} with only small changes due to different boundary conditions for \eqref{phi_equation_0} and \eqref{phi_equation_tilde}.
Using the boundary condition \eqref{boundary_condition_r_2} together with $\tilde{U}(\phi_0)=0$, we find
\begin{align}
  &r(\phi)=R+\int_{\phi_{*}}^\phi d\phi_1\,(2\tilde{U}(\phi_1))^{-1/2}, \label{r_phi}\\
  &R=\int_{\phi_0}^{\phi_{*}} d\phi_1\,(2\tilde{U}(\phi_1))^{-1/2} \label{R_tilde_u}
\end{align}
instead of \eqref{bounce_0_with_R} and \eqref{bounce_0_R}. 
The equations \eqref{r_phi} and \eqref{R_tilde_u} would completely solve the problem of finding $\phi(r)$ for a given $U(\phi)$ if it were not for the need to determine $\tilde{U}(\phi)$ without knowing $\phi(r)$. 
We set out towards an eventual resolution of this difficulty by first examining the relationship between $U$ and $\tilde{U}$ more closely.

Integrating \eqref{u_phi} over $\phi$, substituting $(d\phi/dr)=(2\tilde{U})^{1/2}$ and using the boundary values $U(\phi_{+})=0$ and $\tilde{U}(\phi_{+})=0$, we find
\begin{align}
  U(\phi)=\tilde{U}(\phi)-n\int_{\phi}^{\phi_+} d\phi_1\,(2\tilde{U}(\phi_1))^{1/2}(r(\phi_1))^{-1}.
  \label{u_u_tilde_r}
\end{align}
Now \eqref{r_phi} leads to
\begin{align}
  U(\phi)=\tilde{U}(\phi)-n\int_{\phi}^{\phi_+} d\phi_1\,\bigl(2\tilde{U}(\phi_1)\bigr)^{1/2}\Bigl[R+\int_{\phi_{*}}^{\phi_1} d\phi_2\,(2\tilde{U}(\phi_2))^{-1/2}\Bigr]^{-1},
  \label{u_u_tilde}
\end{align}
which together with \eqref{R_tilde_u} gives $U$ directly in terms of $\tilde{U}$; unfortunately, we need to reverse this procedure and find $\tilde{U}$ in terms of $U$.

One way to arrive at a formula expressing $\tilde{U}$ in terms of $U$ is to use $(d\phi/dr)=(2\tilde{U})^{1/2}$ in \eqref{u_phi} to rewrite it in the form
\begin{align}
  \frac{dU(\phi(r))}{dr}=\frac{d\tilde{U}(\phi(r))}{dr}+\frac{2n}{r}\tilde{U}(\phi(r)).
  \label{u_r}
\end{align}
Equation \eqref{u_r} is a first-order linear differential equation for $\tilde{U}(\phi(r))$ with the general solution
\begin{align}
  \tilde{U}(\phi(r))=r^{-2n}\biggl[C+\int_0^r dr_1\,r_1^{2n}\frac{dU(\phi(r_1))}{dr_1}\biggr],
  \label{u_tilde_constant}
\end{align}
where the integration constant $C$ can be found as follows.
Equations \eqref{u_small_r} and \eqref{phi_small_r_approximation} give
\begin{align}
  \frac{dU(\phi(r))}{dr}\approx\frac{(\phi_0-\phi_{-})^2}{n+1}k_{-}^4 r, \ r\to 0,
  \label{}
\end{align}
substitution of which into \eqref{u_tilde_constant} leads to
\begin{align}
  \tilde{U}(\phi(r))\approx r^{-2n}\biggl[C+\frac{(\phi_0-\phi_{-})^2}{2(n+1)^2}k_{-}^4 r^{2n+2}\biggr], \ r\to 0.
  \label{}
\end{align}
As a result, $\tilde{U}(\phi(0))=\tilde{U}(\phi_0)=0$ now requires $C=0$.
We finally arrive at
\begin{align}
  \tilde{U}(\phi(r))&=r^{-2n}\int_0^r dr_1\,r_1^{2n}\frac{dU(\phi(r_1))}{dr_1}\nn\\
  &=U(\phi(r))-2nr^{-2n}\int_0^r dr_1\,r_1^{2n-1} U(\phi(r_1)),
  \label{u_tilde}
\end{align}
where the second form, obtained from the first form by integration by parts, might be more convenient for calculations.
Although \eqref{u_tilde} appears to express $\tilde{U}$ in terms of $U$, unfortunately, it also requires the function $\phi(r)$, which itself can be found only when $\tilde{U}$ is known; to avoid circular reasoning here, we cannot solve \eqref{phi_equation} for $\phi(r)$ since $\tilde{U}$ in \eqref{u_tilde} is an instrument towards $\phi(r)$ via \eqref{r_phi}.

\subsection{Expansions}\label{}

Returning now to \eqref{u_u_tilde}, we first notice that it together with \eqref{R_tilde_u} directly gives $U$ in terms of $\tilde{U}$, but since our goal is to find the inverse operation, we face a non-linear integral equation for $\tilde{U}$.
Despite its complexity, \eqref{u_u_tilde} is particularly suitable for developing an iterative method for finding approximations of $\tilde{U}$ in terms of $U$.

A naive method of solving \eqref{u_u_tilde} by iterations does not work.
Indeed, if we ignore the second term on the right-hand side of \eqref{u_u_tilde}, we find the zeroth-order approximation $\tilde{U}\approx U$.
Now substituting this approximation into the right-hand side of \eqref{u_u_tilde}, we find the first-order approximation
\begin{align}
  &\tilde{U}(\phi)\approx U(\phi)+nR^{-1}\int_{\phi}^{\phi_+} d\phi_1\,(2U(\phi_1))^{1/2},
  \label{U_approximation}
\end{align}
while \eqref{R_tilde_u} gives in the zeroth-order approximation
\begin{align}
  &R\approx\int_{\phi_0}^{\phi_{*}} d\phi_1\,(2U(\phi_1))^{-1/2}. \label{R_U_approximation}
\end{align}
Since the energy of the analogous classical particle moving in the potential $-U$ is dissipated for the motion from $\phi_0$ to $\phi_{+}$, it follows that $U(\phi_0)<U(\phi_{+})=0$.
Continuity now implies that $U(\phi)<0$ for some $\phi_0<\phi<\phi_{+}$, and for such $\phi$ the square roots in \eqref{U_approximation} and \eqref{R_U_approximation} will be complex-valued, which is not allowed.

Comparing this situation with the thin-wall approximation in Sec.~\ref{section_thin_wall_approximation}, where the even part of $U(\phi)$ appeared, we see the need to introduce the even and odd parts of the potential functions, 
\begin{align}
  U_{\pm}(\phi)&=\tfrac{1}{2}(U(\phi)\pm U(-\phi)), \label{u_plus_minus}
\\
  \tilde{U}_{\pm}(\phi)&=\tfrac{1}{2}(\tilde{U}(\phi)\pm \tilde{U}(-\phi)), \label{u_tilde_plus_minus}
\end{align}
and rewrite \eqref{u_u_tilde} as
\begin{align}
  U_\pm(\phi)=\tilde{U}_\pm(\phi)-\frac{n}{2}\biggl(\int_{\phi}^{\phi_+}\pm\int_{-\phi}^{\phi_+}\biggr)\,d\phi_1\,(2\tilde{U}(\phi_1))^{1/2}\biggl[R+\int_{\phi_{*}}^{\phi_1} d\phi_2\,(2\tilde{U}(\phi_2))^{-1/2}\biggr]^{-1}.
  \label{u_plus_minus}
\end{align}
Since the function $\tilde{U}$ is defined via \eqref{phi_equation_tilde} only on the interval $[\phi_0,\phi_{+}]$, it follows from $\phi_{-}\le \phi_0$ and $\phi_{-}=-\phi_{+}$ that the functions $\tilde{U}_\pm$ are defined via \eqref{u_tilde_plus_minus} only on the interval $[\phi_0,-\phi_0]$ for $\phi_0\le 0$ and cannot be defined at all for $\phi_0>0$.
Despite this, we will extend the domain of the function $\tilde{U}$ to the whole real line by considering the integral equation \eqref{u_plus_minus} for $\tilde{U}_\pm$ as the definition of $\tilde{U}_\pm$.
We will see in the rest of this section that this extension does not lead to problems when finding $\phi$ through $\tilde{U}$ by inverting \eqref{r_phi} since we restrict the domain of the function $r(\phi)$ to $[\phi_0,\phi_{+}]$ to obtain only physically meaningful half-bounce solutions.

In what follows we will separate even and odd functions in \eqref{u_plus_minus} with the help of the identities
\begin{align}
  &\biggl(\int_{\phi}^{\phi_+}+\int_{-\phi}^{\phi_+}\biggr) d\phi_1\,f_{\text{even}}(\phi_1)=\int_{\phi_{-}}^{\phi_{+}}d\phi_1\,f_{\text{even}}(\phi_1), \label{f_even_plus}\\
  &\biggl(\int_{\phi}^{\phi_+}-\int_{-\phi}^{\phi_+}\biggr) d\phi_1\,f_{\text{even}}(\phi_1)=-\int_{-\phi}^\phi d\phi_1\,f_{\text{even}}(\phi_1), \label{f_even_minus}\\
  &\biggl(\int_{\phi}^{\phi_+}+\int_{-\phi}^{\phi_+}\biggr) d\phi_1\,f_{\text{odd}}(\phi_1)=2\int_\phi^{\phi_{+}}d\phi_1\,f_{\text{odd}}(\phi_1), \label{f_odd_plus}\\
  &\biggl(\int_{\phi}^{\phi_+}-\int_{-\phi}^{\phi_+}\biggr) d\phi_1\,f_{\text{odd}}(\phi_1)=0, \label{f_odd_minus}
\end{align}
which hold for any even function $f_{\textrm{even}}$ and any odd function $f_{\textrm{odd}}$.

We choose the energy-density difference between the false and true vacua $\epsilon=U(\phi_{+})-U(\phi_{-})$ as a small positive parameter and set
\begin{align}
  &U_{+}(\phi)=O(1),\\
  &U_{-}(\phi)=O(\epsilon).
  \label{}
\end{align}
To develop a perturbation theory for which the thin-wall approximation is the first term in the expansion in terms of $\epsilon$, we consider power series expansions
\begin{align}
  &\tilde{U}_\pm(\phi)=\sum_{k=0}^\infty\tilde{U}_{\pm,k}(\phi), \  \tilde{U}_{\pm,k}(\phi)=O(\epsilon^{k}), \label{}\\
  &R=\sum_{k=0}^\infty R_k, \ R_k=O(\epsilon^{k-1}). \label{R_k}
\end{align}
The $\epsilon$-dependence of $R_k$ is in accordance with the relation $R=O(\epsilon^{-1})$ in \eqref{R_approximation}.
We note that we will find $R$ iteratively directly from \eqref{u_plus_minus}; in contrast, the thin-wall approximation in Sec. \ref{section_thin_wall_approximation} relied on maximizing the action with respect to $R$.

Once the iterative expansions for $\tilde{U}_\pm(\phi)$ and $R$ are found, we proceed to the corresponding expansions for the bounce solution $\phi(r)$. 
To this end, we first expand the function $r=\rho(\phi)$ into a power series,
\begin{align}
  &\rho(\phi)=\sum_{k=0}^\infty\rho_k(\phi), \ \rho_0(\phi)=O(\epsilon^{-1}), \ \rho_k(\phi)=O(\epsilon^k), \ k\ge 1, \label{rho}
\end{align}
and find each term $\rho_k$ recursively from \eqref{r_phi}.
We can stop here if the bounce solution in terms of the inverse function  $r=\rho(\phi)$ is sufficient for our purposes, but we can also proceed to finding the direct function $\phi=f(r)$ (which is the inverse of the function $r=\rho(\phi)$) iteratively at the cost of slightly reducing the accuracy. 
Namely, we seek $\phi=f(r)$ as a power series
\begin{align}
  &f(r)=\sum_{k=0}^\infty f_k(r), \ f_k(r)=O(\epsilon^k), \ k\ge 0. \label{f}
\end{align}
Here $\phi=f_0(r)$ is the inverse of the function $r=\rho_0(\phi)$ and $f_k$ for $k\ge 1$ are found recursively from the identity $\rho(f(r))=r$.
Since this step depends on finding an analytic expression for $f_0$, it cannot be done for an arbitrary $U$ and we will perform it only for the specific potential function considered in the example in Sec.~\ref{section_polynomial}. 

We make a note on the notation in the proceeding paragraph.
We distinguish the inverse function in the half-bounce solution $r=\rho(\phi)$ from the generic variable $r=r(\phi)$ appearing in the derivation of the half-bounce solution; similarly, $\phi=f(r)$ is the direct function in the half-bounce solution and $\phi=\phi(r)$ is the generic variable. 
The specific functions $\rho(\phi)$ and $f(r)$ will appear again (through their expansions in terms of $\rho_k(\phi)$ and $f_k(r)$) only in the end of this section and in Secs.~\ref{section_polynomial} and \ref{section_discussion}.

\subsection{Orders zero through four}\label{subsection_orders}

Let us return to the iterative solution of \eqref{u_plus_minus}.
Since $U$ and $\tilde{U}$ are nearly equal and the asymmetry of $U$ is small, we need to set $\tilde{U}$ to be an even function in the zeroth order to start the iteration of the approximating sequence,
\begin{align}
  \tilde{U}_{-,0}(\phi)=0.
  \label{u_tilde_minus_0}
\end{align}
To extract  $O(1)$ and $O(\epsilon)$ terms from \eqref{u_plus_minus}, it is enough to keep only the term $R_0=O(\epsilon^{-1})$ in the brackets there. 
Although the resulting factor $R_0^{-1}=O(\epsilon)$ makes any terms in $\tilde{U}(\phi_1)$ smaller than $O(1)$ irrelevant for this approximation order, we nevertheless keep the term $O(\epsilon)$ in $\tilde{U}(\phi_1)$ to avoid the problem of complex-valued square roots as in \eqref{U_approximation} and \eqref{R_U_approximation}.
With these steps, \eqref{u_plus_minus} becomes
\begin{align}
  U_{+}(\phi)&=\tilde{U}_{+,0}(\phi)+\tilde{U}_{+,1}(\phi)-\tfrac{1}{2}nR_{0}^{-1}\int_{\phi_-}^{\phi_+}d\phi_1\,(2\tilde{U}_{+,0}(\phi_1)+2\tilde{U}_{+,1}(\phi_1))^{1/2}+O(\epsilon^2), \label{u_plus_0}\\
  U_{-}(\phi)&=\tilde{U}_{-,1}(\phi)+\tfrac{1}{2}nR_{0}^{-1}\int_{-\phi}^{\phi}d\phi_1\,(2\tilde{U}_{+,0}(\phi_1)+2\tilde{U}_{+,1}(\phi_1))^{1/2}+O(\epsilon^2), \label{u_minus_0}
\end{align}
where we have used \eqref{f_even_plus} and \eqref{f_even_minus}.

The terms $O(1)$ and $O(\epsilon)$ in \eqref{u_plus_0} give
\begin{align}
  &\tilde{U}_{+,0}(\phi)=U_{+}(\phi), \label{u_tilde_plus_0}\\
  &\tilde{U}_{+,1}(\phi)=\tfrac{1}{2}nR_{0}^{-1}\int_{\phi_-}^{\phi_+}d\phi_1\,(2U_{+}(\phi_1)+2\tilde{U}_{+,1}(\phi_1))^{1/2}, \label{u_tilde_plus_1_R_0}
\end{align}
respectively.
The integral equation \eqref{u_tilde_plus_1_R_0} can be trivially solved for $\tilde{U}_{+,1}$.
Indeed, since the right-hand side of \eqref{u_tilde_plus_1_R_0} does not depend on $\phi$, it means that
\begin{align}
  \tilde{U}_{+,1}(\phi)=\delta_1, \ \delta_1=O(\epsilon)
  \label{u_tilde_plus_1_delta_1}
\end{align}
for some infinitesimal constant $\delta_1$, substitution of which into \eqref{u_tilde_plus_1_R_0} gives
\begin{align}
  R_0=\tfrac{1}{2}n\delta_1^{-1}\int_{\phi_-}^{\phi_+}d\phi_1\,(2U_{+}(\phi_1)+2\delta_1)^{1/2}. \label{R_0}
\end{align}
The term $O(\epsilon)$ in \eqref{u_minus_0} gives
\begin{align}
  \tilde{U}_{-,1}(\phi)=U_{-}(\phi)-\tfrac{1}{2}nR_{0}^{-1}\int_{-\phi}^{\phi}d\phi_1\,(2U_{+}(\phi_1)+2\delta_1)^{1/2}. \label{u_tilde_minus_1}
\end{align}

The $O(\epsilon)$ constant $\delta_1$ is arbitrary and a choice for its value effects terms of all orders in our expansions.
We require 
\begin{align}
  \delta_1\ge-\min_{\phi_{-}\le\phi\le\phi_{+}} U_{+}(\phi), \label{delta_1_inequality}
\end{align}
so that the square roots in \eqref{R_0} and \eqref{u_tilde_minus_1} are real-valued.
We find from  \eqref{u_tilde_minus_1} that $\tilde{U}_{-,1}(\phi_{+})=(\epsilon/2)-\delta_1$, which implies that $\tilde{U}(\phi_{+})=O(\epsilon^2)$ holds for any value of $\delta_1$ since
\begin{align}
  \tilde{U}_{+,0}(\phi_{+})+\tilde{U}_{-,0}(\phi_{+})+\tilde{U}_{+,1}(\phi_{+})+\tilde{U}_{-,1}(\phi_{+})=0.
  \label{u_tilde_phi_plus}
\end{align}
As we have seen in Sec.~\ref{section_bounces}, no generality is lost upon choosing $U$ to satisfy \eqref{u_bound_epsilon}, so that \eqref{delta_1_inequality} becomes $\delta_1\ge\epsilon/2$. 
From now on we set the value of $\delta_1$ to its lower bound,
\begin{align}
  \delta_1=\frac{\epsilon}{2}. \label{delta_1}
\end{align}
One of the reasons for this choice is that now $\tilde{U}_{-,1}(\phi_{+})=0$, which is analogous to $\tilde{U}_{-,0}(\phi_{+})=0$ (although, more generally, $\tilde{U}_{-,0}(\phi)=0$ for any $\phi$).

Equations \eqref{u_tilde_minus_0}, \eqref{u_tilde_plus_0}, \eqref{u_tilde_plus_1_delta_1}, \eqref{R_0}, \eqref{u_tilde_minus_1} and \eqref{delta_1} give the zeroth-order and first-order approximations, which coincide with the thin-wall approximation.

To proceed to terms $O(\epsilon^k)$ in \eqref{u_plus_minus}, we expand $(2\tilde{U}(\phi_1))^{1/2}$ to terms $O(\epsilon^{k-1})$, $R$ to terms $O(\epsilon^{k-2})$ and $(2\tilde{U}(\phi_2))^{-1/2}$ to terms $O(\epsilon^{k-2})$.
Let us work through the cases with $2\le k\le 4$, for which we need the expansion
\begin{align}
  &U_\pm(\phi)=\tilde{U}_{\pm,0}(\phi)+\tilde{U}_{\pm,1}(\phi)+\tilde{U}_{\pm,2}(\phi)+\tilde{U}_{\pm,3}(\phi)+\tilde{U}_{\pm,4}(\phi)-\frac{n}{2}\biggl(\int_{\phi}^{\phi_+}\pm\int_{-\phi}^{\phi_+}\biggr)\, d\phi_1\,p(\phi_1)\nn\\
  &\times\biggl[1+\frac{\tilde{U}_{-,1}(\phi_1)}{p(\phi_1)^2}-\frac{\tilde{U}_{-,1}(\phi_1)^2}{2p(\phi_1)^4}+\frac{\tilde{U}_{+,2}(\phi_1)+\tilde{U}_{-,2}(\phi_1)}{p(\phi_1)^2}+\frac{\tilde{U}_{-,1}(\phi_1)^3}{2p(\phi_1)^6}\nn\\
  &-\frac{\tilde{U}_{-,1}(\phi_1)(\tilde{U}_{+,2}(\phi_1)+\tilde{U}_{-,2}(\phi_1))}{p(\phi_1)^4}+\frac{\tilde{U}_{+,3}(\phi_1)+\tilde{U}_{-,3}(\phi_1)}{p(\phi_1)^2}+O(\epsilon^4)\biggr]\nn\\
  &\times\frac{1}{R_0}\biggl[1-\frac{1}{R_0}\biggl(R_1+\int_{\phi_*}^{\phi_1}\frac{d\phi_2}{p(\phi_2)}\biggr)+\frac{1}{R_0^2}\biggl(R_1+\int_{\phi_*}^{\phi_1}\frac{d\phi_2}{p(\phi_2)}\biggr)^2\nn\\
  &-\frac{1}{R_0}\biggl(R_2-\int_{\phi_*}^{\phi_1} d\phi_2\,\frac{\tilde{U}_{-,1}(\phi_2)}{p(\phi_2)^3}\biggr)-\frac{1}{R_0^3}\biggl(R_1+\int_{\phi_*}^{\phi_1}\frac{d\phi_2}{p(\phi_2)}\biggr)^3\nn\\
  &+\frac{2}{R_0^2}\biggl(R_1+\int_{\phi_*}^{\phi_1}\frac{d\phi_2}{p(\phi_2)}\biggr)\biggl(R_2-\int_{\phi_*}^{\phi_1} d\phi_2\,\frac{\tilde{U}_{-,1}(\phi_2)}{p(\phi_2)^3}\biggr)\nn\\
&-\frac{1}{R_0}\biggl(R_3+\int_{\phi_*}^{\phi_1} d\phi_2\biggl(\frac{3\tilde{U}_{-,1}(\phi_2)^2}{2p(\phi_2)^5}-\frac{\tilde{U}_{+,2}(\phi_2)+\tilde{U}_{-,2}(\phi_2)}{p(\phi_2)^3}\biggr)\biggr)+O(\epsilon^3)\biggr],
  \label{u_plus_minus_order_2_and_3}
\end{align}
where
\begin{align}
  p(\phi)=(2U_{+}(\phi)+\epsilon)^{1/2}.
  \label{p}
\end{align}
Using \eqref{f_even_plus}, \eqref{f_even_minus}, \eqref{f_odd_plus}, \eqref{f_odd_minus}, \eqref{u_tilde_minus_0}, \eqref{u_tilde_plus_0}, \eqref{u_tilde_plus_1_delta_1}, \eqref{R_0}, \eqref{u_tilde_minus_1} and \eqref{delta_1} for terms $O(\epsilon^2)$ in \eqref{u_plus_minus_order_2_and_3}, we find
\begin{align}
  &\tilde{U}_{+,2}(\phi)=-\frac{\epsilon}{2R_0}\biggl(R_1-\int_0^{\phi_*}\frac{d\phi_1} {p(\phi_1)}\biggr)+\frac{n}{R_0}\int_\phi^{\phi_{+}}d\phi_1\,\frac{\tilde{U}_{-,1}(\phi_1)}{p(\phi_1)}\nn\\
  &-\frac{n}{R_0^2}\int_\phi^{\phi_{+}}d\phi_1\, p(\phi_1)\int_0^{\phi_1}\frac{d\phi_2}{p(\phi_2)}, \label{u_tilde_plus_2_R_1}\\
  &\tilde{U}_{-,2}(\phi)=\frac{n}{2R_0^2}\biggl(R_1-\int_0^{\phi_*}\frac{d\phi_1}{p(\phi_1)}\biggr)\int_{-\phi}^\phi d\phi_2\,p(\phi_2). \label{u_tilde_minus_2_R_1}
\end{align}
Note that $\tilde{U}(\phi_{+})=O(\epsilon^3)$ holds for any value of $R_1$ due to \eqref{u_tilde_phi_plus} and 
\begin{align}
  &\tilde{U}_{+,2}(\phi_{+})+\tilde{U}_{-,2}(\phi_{+})=0.
  \label{}
\end{align}

The quantity $R_1$ is undetermined.
However, if we follow the previously derived $\tilde{U}_{-,0}(\phi_{+})=0$ and $\tilde{U}_{-,1}(\phi_{+})=0$ with the analogous $\tilde{U}_{-,2}(\phi_{+})=0$, we find
\begin{align}
  R_1=\int_0^{\phi_*}\frac{d\phi}{p(\phi)},
  \label{R_1}
\end{align}
which we set from now on.
As a result,
\begin{align}
  &\tilde{U}_{+,2}(\phi)=\frac{n}{R_0}\int_\phi^{\phi_{+}}d\phi_1\,\frac{\tilde{U}_{-,1}(\phi_1)}{p(\phi_1)}-\frac{n}{R_0^2}\int_\phi^{\phi_{+}}d\phi_1\, p(\phi_1)\int_0^{\phi_1}\frac{d\phi_2}{p(\phi_2)}, \label{u_tilde_plus_2}\\
  &\tilde{U}_{-,2}(\phi)=0.
  \label{u_tilde_minus_2}
\end{align}
Note that $\tilde{U}_{-,2}(\phi)=0$ is analogous to $\tilde{U}_{-,0}(\phi)=0$ found earlier.

Equations \eqref{R_1}, \eqref{u_tilde_plus_2} and \eqref{u_tilde_minus_2} give the second-order approximation.

Using \eqref{f_even_plus}, \eqref{f_even_minus}, \eqref{f_odd_plus}, \eqref{f_odd_minus}, \eqref{u_tilde_minus_0}, \eqref{u_tilde_plus_0}, \eqref{u_tilde_plus_1_delta_1}, \eqref{R_0}, \eqref{u_tilde_minus_1}, \eqref{delta_1}, \eqref{R_1}, \eqref{u_tilde_plus_2} and \eqref{u_tilde_minus_2} for terms $O(\epsilon^3)$ in \eqref{u_plus_minus_order_2_and_3}, we find
\begin{align}
  &\tilde{U}_{+,3}(\phi)=\frac{n}{2R_0}\int_{\phi_{-}}^{\phi_{+}} d\phi_1\,p(\phi_1)\biggl[-\frac{R_2}{R_0}+\frac{1}{R_0^2}\biggl(\int_{0}^{\phi_1} \frac{d\phi_2}{p(\phi_2)}\biggr)^2-\frac{\tilde{U}_{-,1}(\phi_1)^2}{2p(\phi_1)^4}+\frac{\tilde{U}_{+,2}(\phi_1)}{p(\phi_1)^2}\nn\\
  &-\frac{\tilde{U}_{-,1}(\phi_1)}{R_0 p(\phi_1)^2}\int_0^{\phi_1} \frac{d\phi_2}{p(\phi_2)}+\frac{1}{R_0}\int_{\phi_*}^{\phi_1} d\phi_2\,\frac{\tilde{U}_{-,1}(\phi_2)}{p(\phi_2)^3}\biggr], \label{u_tilde_plus_3_with_delta_3}\\
  &\tilde{U}_{-,3}(\phi)=-\frac{n}{2R_0}\int_{-\phi}^{\phi} d\phi_1\,p(\phi_1)\biggl[-\frac{R_2}{R_0}+\frac{1}{R_0^2}\biggl(\int_{0}^{\phi_1} \frac{d\phi_2}{p(\phi_2)}\biggr)^2-\frac{\tilde{U}_{-,1}(\phi_1)^2}{2p(\phi_1)^4}+\frac{\tilde{U}_{+,2}(\phi_1)}{p(\phi_1)^2}\nn\\
  &-\frac{\tilde{U}_{-,1}(\phi_1)}{R_0 p(\phi_1)^2}\int_0^{\phi_1} \frac{d\phi_2}{p(\phi_2)}+\frac{1}{R_0}\int_{\phi_*}^{\phi_1} d\phi_2\,\frac{\tilde{U}_{-,1}(\phi_2)}{p(\phi_2)^3}\biggr]. \label{u_tilde_minus_3}
\end{align}
Since the right-hand side of \eqref{u_tilde_plus_3_with_delta_3} does not depend on $\phi$, it follows that
\begin{align}
  \tilde{U}_{+,3}(\phi)=\delta_3, \ \delta_3=O(\epsilon^3)
  \label{u_tilde_plus_3_delta_3}
\end{align}
for some infinitesimal constant $\delta_3$.

The quantities $R_2$ and $\delta_3$ are undetermined.
Similar to $\tilde{U}_{-,k}(\phi_{+})=0$ for $0\le k\le 2$, we set $\tilde{U}_{-,3}(\phi_{+})=0$, and solve \eqref{u_tilde_plus_3_with_delta_3}, \eqref{u_tilde_minus_3} and \eqref{u_tilde_plus_3_delta_3} for $R_2$ and $\delta_3$ to find
\begin{align}
  &R_2=\frac{n}{\epsilon}\int_{\phi_{-}}^{\phi_{+}} d\phi_1\,p(\phi_1)\biggl[\frac{1}{R_0^2}\biggl(\int_{0}^{\phi_1}\frac{d\phi_2}{p(\phi_2)}\biggr)^2-\frac{\tilde{U}_{-,1}(\phi_1)^2}{2p(\phi_1)^4}+\frac{\tilde{U}_{+,2}(\phi_1)}{p(\phi_1)^2}\nn\\
  &-\frac{\tilde{U}_{-,1}(\phi_1)}{R_0 p(\phi_1)^2}\int_0^{\phi_1} \frac{d\phi_2}{p(\phi_2)}+\frac{1}{R_0}\int_{\phi_*}^{\phi_1} d\phi_2\,\frac{\tilde{U}_{-,1}(\phi_2)}{p(\phi_2)^3}\biggr],\label{R_2}\\
  &\delta_3=0. \label{delta_3}
\end{align}

Equations \eqref{u_tilde_minus_3}, \eqref{u_tilde_plus_3_delta_3}, \eqref{R_2} and \eqref{delta_3} give the third-order approximation.

Using \eqref{f_even_plus}, \eqref{f_even_minus}, \eqref{f_odd_plus}, \eqref{f_odd_minus}, \eqref{u_tilde_minus_0}, \eqref{u_tilde_plus_0}, \eqref{u_tilde_plus_1_delta_1}, \eqref{R_0}, \eqref{u_tilde_minus_1}, \eqref{delta_1}, \eqref{R_1}, \eqref{u_tilde_plus_2}, \eqref{u_tilde_minus_2}, \eqref{u_tilde_minus_3}, \eqref{u_tilde_plus_3_delta_3}, \eqref{R_2} and \eqref{delta_3} for terms $O(\epsilon^4)$ in \eqref{u_plus_minus_order_2_and_3}, we find
\begin{align}
  &\tilde{U}_{+,4}(\phi)=\frac{\epsilon}{2R_0}\biggl[-R_3+\int_0^{\phi_*}d\phi_2\biggl(\frac{3\tilde{U}_{-,1}(\phi_2)^2}{2p(\phi_2)^5}-\frac{\tilde{U}_{+,2}(\phi_2)}{p(\phi_2)^3}\biggr)\biggr]+\frac{n}{2R_0^4}\int_\phi^{\phi_{+}}d\phi_1\,p(\phi_1)\nn\\
  &\times\biggl[R_0^3\biggl(\frac{\tilde{U}_{-,1}(\phi_{1})^3}{2p(\phi_1)^6}-\frac{\tilde{U}_{-,1}(\phi_1)\tilde{U}_{+,2}(\phi_1)}{p(\phi_1)^4}+\frac{\tilde{U}_{-,3}(\phi_1)}{p(\phi_1)^2}\biggr)-\biggl(\int_0^{\phi_1}\frac{d\phi_2}{p(\phi_2)}\biggr)^3\nn\\
  &+R_0\frac{\tilde{U}_{-,1}(\phi_1)}{p(\phi_1)^2}\biggl(\int_0^{\phi_1}\frac{d\phi_2}{p(\phi_2)}\biggr)^2+R_0^2\biggl(\frac{\tilde{U}_{-,1}(\phi_1)^2}{2p(\phi_1)^4}-\frac{\tilde{U}_{+,2}(\phi_1)}{p(\phi_1)^2}\biggr)\int_0^{\phi_1}\frac{d\phi_2}{p(\phi_2)}\nn\\
&-R_0^2\int_0^{\phi_1}d\phi_2\biggl(\frac{3\tilde{U}_{-,1}(\phi_2)^2}{2p(\phi_2)^5}-\frac{\tilde{U}_{+,2}(\phi_2)}{p(\phi_2)^3}\biggr)+R_0\biggl(2\int_0^{\phi_1}\frac{d\phi_2}{p(\phi_2)}-R_0\frac{\tilde{U}_{-,1}(\phi_1)}{p(\phi_1)^2}\biggr)\nn\\
&\times\biggl(R_2-\int_{\phi_{*}}^{\phi_1}d\phi_2\,\frac{\tilde{U}_{-,1}(\phi_2)}{p(\phi_2)^3}\biggr)\biggr],\label{u_tilde_plus_4_with_R_3}\\
  &\tilde{U}_{-,4}(\phi)=\frac{n}{2R_0^2}\int_{-\phi}^\phi d\phi_1\,p(\phi_1)\biggl[R_3+\int_0^{\phi_*}d\phi_2\biggl(-\frac{3\tilde{U}_{-,1}(\phi_2)^2}{2p(\phi_2)^5}+\frac{\tilde{U}_{+,2}(\phi_2)}{p(\phi_2)^3}\biggr)\biggr].\label{u_tilde_minus_4_with_R_3}
\end{align}
The quantity $R_3$ is undetermined.
Similar to $\tilde{U}_{-,k}(\phi_{+})=0$ for $0\le k\le 3$, we set $\tilde{U}_{-,4}(\phi_{+})=0$ and solve \eqref{u_tilde_minus_4_with_R_3} for $R_3$ to find
\begin{align}
  R_3=\int_0^{\phi_*}d\phi\biggl(\frac{3\tilde{U}_{-,1}(\phi)^2}{2p(\phi)^5}-\frac{\tilde{U}_{+,2}(\phi)}{p(\phi)^3}\biggr),\label{R_3}
\end{align}
which leads to
\begin{align}
  &\tilde{U}_{+,4}(\phi)=\frac{n}{2R_0^4}\int_\phi^{\phi_{+}}d\phi_1\,p(\phi_1)\biggl[R_0^3\biggl(\frac{\tilde{U}_{-,1}(\phi_1)^3}{2p(\phi_1)^6}-\frac{\tilde{U}_{-,1}(\phi_1)\tilde{U}_{+,2}(\phi_1)}{p(\phi_1)^4}+\frac{\tilde{U}_{-,3}(\phi_1)}{p(\phi_1)^2}\biggr)\nn\\
    &-\biggl(\int_0^{\phi_1}\frac{d\phi_2}{p(\phi_2)}\biggr)^3+R_0\frac{\tilde{U}_{-,1}(\phi_1)}{p(\phi_1)^2}\biggl(\int_0^{\phi_1}\frac{d\phi_2}{p(\phi_2)}\biggr)^2+R_0^2\biggl(\frac{\tilde{U}_{-,1}(\phi_1)^2}{2p(\phi_1)^4}-\frac{\tilde{U}_{+,2}(\phi_1)}{p(\phi_1)^2}\biggr)\int_0^{\phi_1}\frac{d\phi_2}{p(\phi_2)}\nn\\
&-R_0^2\int_0^{\phi_1}d\phi_2\biggl(\frac{3\tilde{U}_{-,1}(\phi_2)^2}{2p(\phi_2)^5}-\frac{\tilde{U}_{+,2}(\phi_2)}{p(\phi_2)^3}\biggr)+R_0\biggl(2\int_0^{\phi_1}\frac{d\phi_2}{p(\phi_2)}-R_0\frac{\tilde{U}_{-,1}(\phi_1)}{p(\phi_1)^2}\biggr)\nn\\
&\times\biggl(R_2-\int_{\phi_{*}}^{\phi_1}d\phi_2\,\frac{\tilde{U}_{-,1}(\phi_2)}{p(\phi_2)^3}\biggr)\biggr],\label{u_tilde_plus_4}\\
  &\tilde{U}_{-,4}(\phi)=0.\label{u_tilde_minus_4}
\end{align}
Note that $\tilde{U}_{-,4}(\phi)=0$ is analogous to $\tilde{U}_{-,0}(\phi)=0$ and $\tilde{U}_{-,2}(\phi)=0$ found earlier.

Equations \eqref{R_3}, \eqref{u_tilde_plus_4} and \eqref{u_tilde_minus_4} give the fourth-order approximation.

To obtain approximations for the half-bounce solution, we expand the right-hand side of \eqref{r_phi} to terms $O(\epsilon^3)$ and find the first few terms in \eqref{rho},
\begin{align}
  &\rho_0(\phi)=R_0+R_1+\int_{\phi_*}^\phi\frac{d\phi_1}{p(\phi_1)},\label{rho_0}\\
  &\rho_1(\phi)=R_2-\int_{\phi_*}^\phi d\phi_1\frac{\tilde{U}_{-,1}(\phi_1)}{p(\phi_1)^3},\label{rho_1}\\
  &\rho_2(\phi)=R_3+\int_{\phi_*}^\phi d\phi_1\biggl(\frac{3\tilde{U}_{-,1}(\phi_1)^2}{2p(\phi_1)^5}-\frac{\tilde{U}_{+,2}(\phi_1)}{p(\phi_1)^3}\biggr).\label{rho_2}
\end{align}
The first few terms in \eqref{f} are found similarly.
After solving
\begin{align}
  &r=R_0+R_1+\int_{\phi_*}^{f_0(r)}\frac{d\phi_1}{p(\phi_1)}\label{psi_0}
\end{align}
for the function $f_0(r)$, we iterate and find
\begin{align}
  &f_1(r)=p(f_0(r))\biggl(-R_2+\int_{\phi_*}^{f_0(r)}d\phi_1\frac{\tilde{U}_{-,1}(\phi_1)}{p(\phi_1)^3}\biggr),\label{psi_1}\\
  &f_2(r)=p(f_0(r))\biggl[\frac{f_1(r)^2}{2p(f_0(r))^2}\frac{dp(f_0(r))}{df_0(r)}+\frac{\tilde{U}_{-,1}(f_0(r))}{p(f_0(r))^3}f_1(r)\nn\\
  &-\int_0^{f_0(r)} d\phi_1\biggl(\frac{3\tilde{U}_{-,1}(\phi_1)^2}{2p(\phi_1)^5}-\frac{\tilde{U}_{+,2}(\phi_1)}{p(\phi_1)^3}\biggr)\biggr].\label{psi_2}
\end{align}

It is clear how to proceed to higher orders.

\section{Fourth-order polynomial potentials}\label{section_polynomial}

Although the approximation method developed in Sec.~\ref{section_method} works for an arbitrary continuously differential function $U$ with two minima and one maximum, it may appear that the formulas derived for various orders of approximation are difficult to use in practice.
In particular, one may expect the need for many terms in the approximation in order to achieve an acceptable accuracy for a half-bounce with the thick wall, or that the higher order approximations will become too complicated to be useful.

To demonstrate applicability of the above results, we consider in this section an example of the general fourth-order polynomial potential
\begin{align}
  U(\phi)=a_0+a_1\phi+a_2\phi^2+a_3\phi^3+a_4\phi^4.
  \label{}
\end{align}
We now proceed to deduce the values of the coefficients $\{a_k\}$ by imposing the conditions specified in Sec.~\ref{section_bounces} on the function $U$, which have been shown to lead to no loss of generality.
We first choose $\phi_{-}=-1$ and $\phi_{+}=1$, which give $a_3=-\frac{1}{3}a_1$ and $a_4=-\frac{1}{2}a_2$.
The requirement $U(1)=0$ now leads to $a_2=-2a_0-\frac{4}{3}a_1$, after which $U(-1)=-\epsilon$ yields $a_1=\frac{3}{4}\epsilon$, which result in
\begin{align}
  a_1=\frac{3}{4}\epsilon,\ a_2=-2a_0-\epsilon,\ a_3=-\frac{1}{4}\epsilon,\ a_4=a_0+\frac{1}{2}\epsilon.
  \label{a_polynomial}
\end{align}
For the function $U$ to have two minima and one maximum, we need $a_4>0$, which implies $a_0>-\frac{1}{2}\epsilon$ and $a_2<0$.
We compute
\begin{align}
  U(\phi)+U(-\phi)=(2a_0+\epsilon)(1-\phi^2)^2-\epsilon\ge -\epsilon
  \label{}
\end{align}
and find that \eqref{u_bound_epsilon} is satisfied.
We thus have freedom to impose one additional condition on the function $U$, for which we choose $a_0=(1-\epsilon)/2$ without any loss of generality.
As a result,
\begin{align}
  &a_0=\frac{1-\epsilon}{2}, \ a_1=\frac{3\epsilon}{4},\ a_2=-1,\ a_3=-\frac{\epsilon}{4},\ a_4=\frac{1}{2},\\
  &U(\phi)=\frac{1}{2}(1-\phi^2)^2-\frac{\epsilon}{2}+\frac{\epsilon\phi}{4}(3-\phi^2). \label{polynomial_u}
\end{align}
We find
\begin{align}
  \phi_*=\frac{3\epsilon}{8}
  \label{phi_star}
\end{align}
and see that we need $0<\epsilon<\frac{8}{3}$ in order for $\phi=1$ to be the relative minimum. 

Using equations from Sec.~\ref{section_method}, we find all the results necessary to obtain the approximations for the half-bounce for this example through the fourth order.
In the remainder of this section, all functions of $\phi$ will be restricted to the domain $(\phi_{-},\phi_{+})=(-1,1)$.

We start with
\begin{align}
  &p(\phi)=1-\phi^2
\end{align}
and use \eqref{u_tilde_minus_0}, \eqref{u_tilde_plus_0}, \eqref{u_tilde_plus_1_delta_1}, \eqref{R_0}, \eqref{u_tilde_minus_1} and \eqref{delta_1} to find the zeroth-order and first-order quantities
\begin{align}
  &\tilde{U}_{-,0}(\phi)=0,\\
  &\tilde{U}_{+,0}(\phi)=\frac{1}{2}((1-\phi^2)^2-\epsilon),\\
  &\tilde{U}_{+,1}(\phi)=\frac{\epsilon}{2},\\
  &R_0=\frac{4n}{3\epsilon},\\
  &\tilde{U}_{-,1}(\phi)=0.
\end{align}
Up to this order, our approximation coincides with the thin-wall approximation.
Going beyond it, \eqref{R_1}, \eqref{u_tilde_plus_2} and \eqref{u_tilde_minus_2} give the second-order quantities
\begin{align}
  &R_1=\frac{1}{2}\ln{\frac{8+3\epsilon}{8-3\epsilon}},\label{R_1_polynomial}\\
  &\tilde{U}_{+,2}=\frac{3\epsilon^2}{32n}\Bigl(\phi(3-\phi^2)\ln{\frac{1+\phi}{1-\phi}}+2\ln{\frac{1-\phi^2}{4}}+1-\phi^2\Bigr),\\
  &\tilde{U}_{-,2}(\phi)=0.
\end{align}
Now \eqref{u_tilde_minus_3}, \eqref{u_tilde_plus_3_delta_3}, \eqref{R_2} and \eqref{delta_3} lead to the third-order quantities
\begin{align}
  &\tilde{U}_{+,3}(\phi)=0,\\
  &\tilde{U}_{-,3}(\phi)=\frac{3\epsilon^3}{256n^2}\biggl[\Bigl(3\phi(\phi^2-3)\ln{\frac{1+\phi}{1-\phi}}-6\ln{\frac{1-\phi^2}{4}}+3(n-2)(1-\phi^2)\Bigr)\ln{\frac{1+\phi}{1-\phi}}\nn\\
  &+(n-1)\Bigl(12\li2{\Bigl(\frac{1+\phi}{2}\Bigr)}-12\li2{\Bigl(\frac{1-\phi}{2}\Bigr)}+\phi\bigl((\pi^2-6)\phi^2-3(\pi^2-2)\bigr)\Bigr)\biggr],\\
  &R_2=\frac{(n-1)(6-\pi^2)\epsilon}{16n},
\end{align}
where $\li2$ is the dilogarithm function.
We give the expression for only one fourth-order quantity,
\begin{align}
  &R_3=\frac{3\epsilon^2}{512n}\biggl[6\li2{\Bigl(\frac{8+3\epsilon}{16}\Bigr)}-6\li2{\Bigl(\frac{8-3\epsilon}{16}\Bigr)}-3\ln{\frac{64-9\epsilon^2}{256}}\ln{\frac{8+3\epsilon}{8-3\epsilon}}\nn\\
  &-\frac{6(4096+1152\epsilon^2-135\epsilon^4)}{(64-9\epsilon^2)^2}\ln{\frac{8+3\epsilon}{8-3\epsilon}}-\frac{96\epsilon}{(64-9\epsilon^2)^2}\Bigl(64(1-40\ln{2})\nn\\
&+9\epsilon^2(24\ln{2}-1)+(320-27\epsilon^2)\ln{(64-9\epsilon^2)}\Bigr)\biggr]\nn\\
&=\frac{9(4\ln{2}-1)\epsilon^3}{256n}+O(\epsilon^5),\label{R_3_polynomial}
\end{align}
which we calculate from \eqref{R_3}.

We note that $R_1=O(\epsilon)$ in \eqref{R_1_polynomial} and $R_3=O(\epsilon^3)$ in \eqref{R_3_polynomial} contradict $R_k=O(\epsilon^{k-1})$ in \eqref{R_k}.
Our choice $a_0=(1-\epsilon)/2$ explains this discrepancy because it leads to $\phi_{*}=O(\epsilon)$ in \eqref{phi_star} in contrast to $\phi_{*}=O(1)$ that was assumed in the derivation of approximations in Sec.~\ref{subsection_orders}.
We made such a choice for $a_0$ (which is the only one possible to get $\phi_{*}=O(\epsilon)$) specifically to test accuracy of the iterative method in the worst possible case (for any fourth-order polynomial potential) when the orders of some terms in expansions have slightly different $\epsilon$-dependencies.
Despite these discrepancies, the agreement between exact numerical solutions and approximate analytic solutions is outstanding (see below), and it should be clear that the agreement will only improve for any other choice of $a_0$ satisfying $a_0>-\frac{1}{2}\epsilon$ (as required by \eqref{a_polynomial}).

With these preliminary results, we now obtain from \eqref{rho_0}, \eqref{rho_1} and \eqref{rho_2}

\begin{align}
  &\rho_0(r)=R_0+\frac{1}{2}\ln{\frac{1+\phi}{1-\phi}},\\
  &\rho_1(r)=R_2,\\
  &\rho_2(r)=\frac{3\epsilon^2}{512n}\biggl[6\li2{\Bigl(\frac{1+\phi}{2}\Bigr)}-6\li2{\Bigl(\frac{1-\phi}{2}\Bigr)}-3\ln{\frac{1-\phi^2}{4}}\ln{\frac{1+\phi}{1-\phi}}\nn\\
    &+\frac{1}{(1-\phi^2)^2}\Bigl(2(5\phi^4-6\phi^2-3)\ln{\frac{1+\phi}{1-\phi}}+4\phi(3\phi^2-5)\ln{\frac{1-\phi^2}{4}}-4\phi(1-\phi^2)\Bigr)\biggr],
\end{align}
which via \eqref{psi_0}, \eqref{psi_1} and \eqref{psi_2} finally leads to
approximations for the half-bounce solution
\begin{align}
  &f_0(r)=\tanh{(r-R_0)},\\
  &f_1(r)=-\frac{R_2}{(\cosh{(r-R_0)})^2},\\
  &f_2(r)=-\frac{R_2^2\tanh{(r-R_0)}}{(\cosh{(r-R_0))^2}}+\frac{3\epsilon^2}{256n(\cosh{(r-R_0))^2}}\biggl[3\li2{\Bigl(\frac{1-\tanh{(r-R_0)}}{2}\Bigr)}\nn\\
    &-3\li2{\Bigl(\frac{1+\tanh{(r-R_0)}}{2}\Bigr)}+(r-R_0)\Bigl(-3+8\cosh{2(r-R_0)+\cosh{4(r-R_0)}}\nn\\&-6\ln{(2\cosh{(r-R_0)})}\Bigr)-\ln{(2\cosh{(r-R_0)})}\bigl(8\sinh{2(r-R_0)}+\sinh{4(r-R_0)}\bigr)\nn\\
  &+\sinh{2(r-R_0)}\biggr].
\end{align}

We note that despite complicated intermidiate results leading to the bounce solution $f$, the expressions for the first approximation $f\approx f_0+f_1$ and even the second approximation $f\approx f_0+f_1+f_2$ to some extend are rather simple.
Together with the high numerical accuracy shown in the following section, we view this simplicity as a demonstration of the strength of our approximation method.

\begin{figure}[h]
  \begin{center}
    \includegraphics[width=468pt]{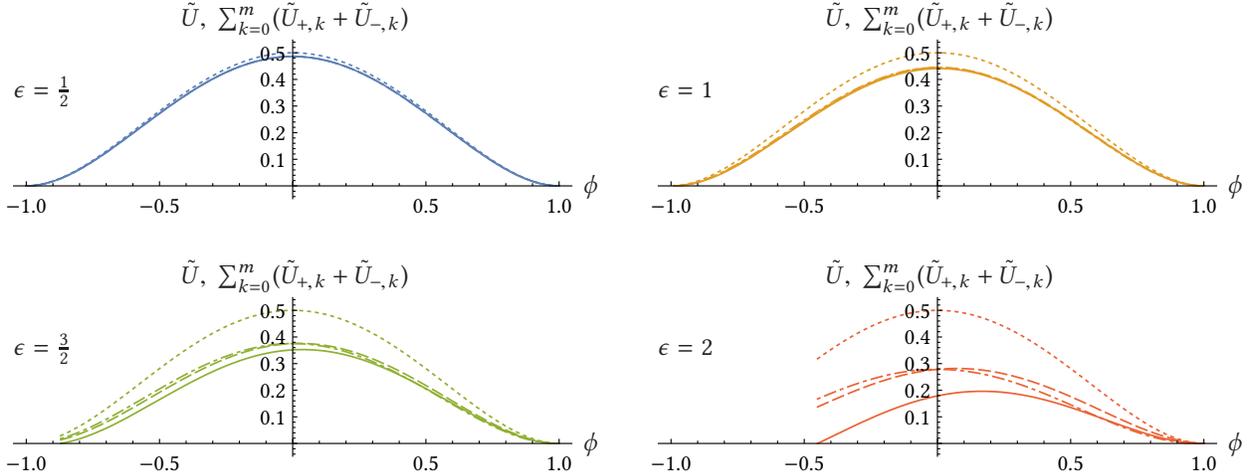}
  \end{center}
  \caption{The exact numerical solution $\tilde{U}$ (the solid curve) and the approximate analytic solutions $\sum_{k=0}^m(\tilde{U}_{+,k}+\tilde{U}_{-,k})$ for $m=1$ (the dotted curve), $m=2$ (the dot-dashed curve) and $m=3$ (the dashed curve) for $n=3$ and the potential \eqref{polynomial_u} with different $\epsilon$.
(Some curves nearly coincide.)
  Each color of curves in this figure and in Figs.~\ref{figure_u_phi} and \ref{figure_phi_0} represents the same value of $\epsilon$. 
}
  \label{figure_u_tilde_0}
\end{figure}

\begin{figure}[h]
  \begin{center}
    \includegraphics[width=468pt]{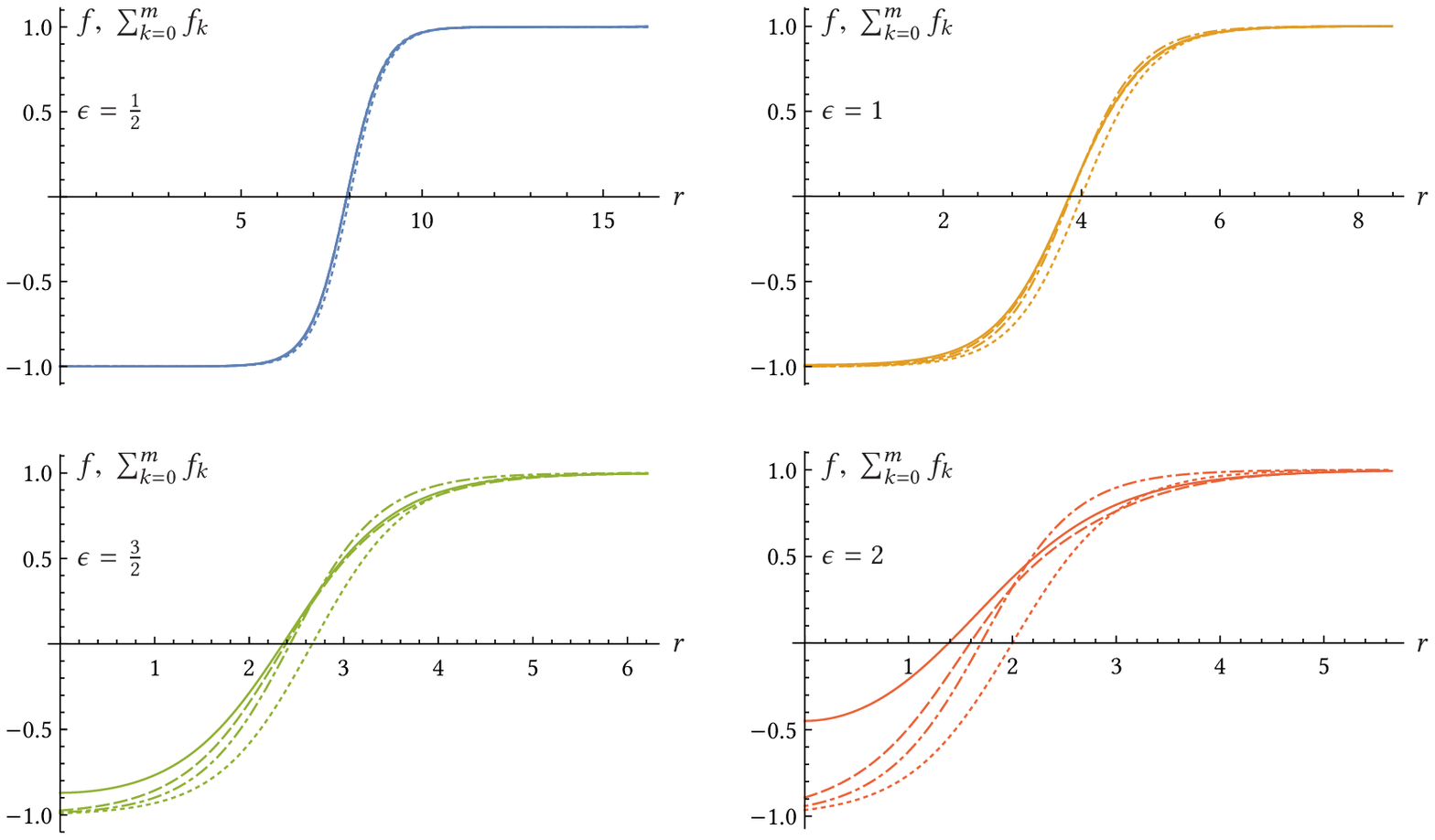}
  \end{center}
  \caption{The exact numerical solution $f$ (the solid curve) and the approximate analytic solutions $\sum_{k=0}^m f_k$ for $m=0$ (the dotted curve), $m=1$ (the dot-dashed curve) and $m=2$ (the dashed curve) for $n=3$ and the potential \eqref{polynomial_u} with different $\epsilon$.
(Some curves nearly coincide.)
  Each color of curves in this figure and in Figs.~\ref{figure_u_phi} and \ref{figure_u_tilde_0} represents the same value of $\epsilon$. 
}
  \label{figure_phi_0}
\end{figure}

\begin{figure}[h]
  \begin{center}
    \includegraphics[width=468pt]{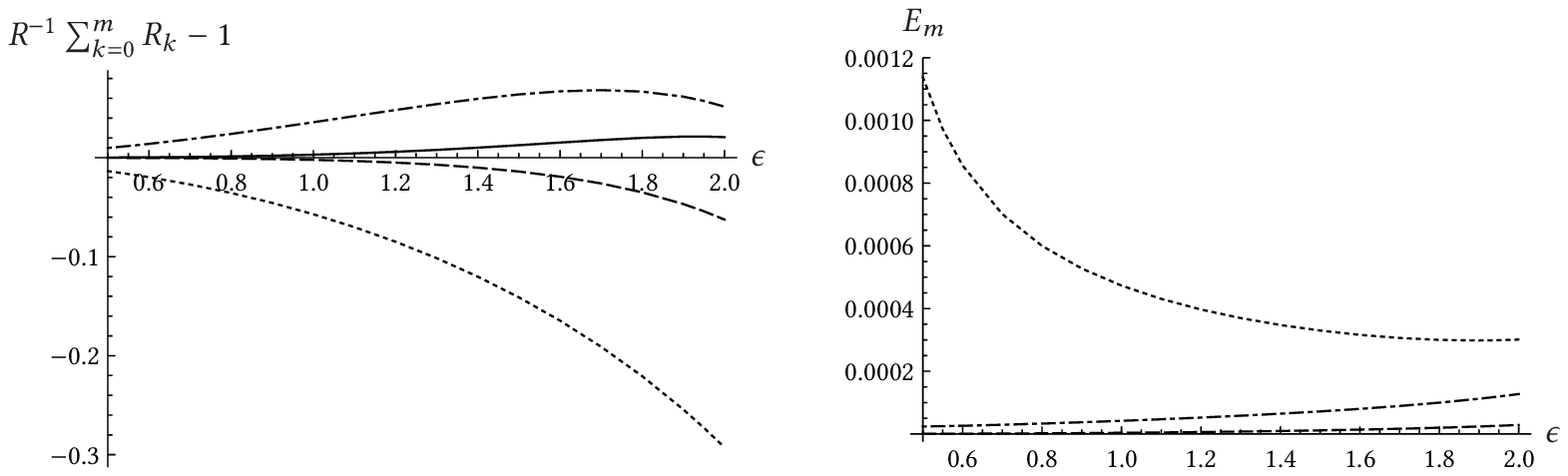}
  \end{center}
  \caption{Left: the $\epsilon$-dependence of $R^{-1}\sum_{k=0}^m R_k-1$ for $m=0$ (the dotted curve), $m=1$ (the dot-dashed curve), $m=2$ (the dashed curve) and $m=3$ (the solid curve) for $n=3$ and the potential \eqref{polynomial_u}.
Right: the $\epsilon$-dependence of $E_m$ for $m=0$ (the dotted curve), $m=1$ (the dot-dashed curve) and $m=2$ (the dashed curve) in \eqref{f_error} for $n=3$ and the potential \eqref{polynomial_u}.}
  \label{figure_r_error_phi_error}
\end{figure}

The accuracy of our successive approximations can be seen in Figs.~\ref{figure_u_tilde_0}, \ref{figure_phi_0} and \ref{figure_r_error_phi_error}. 
In Fig.~\ref{figure_u_tilde_0} we show the exact numerical solution $\tilde{U}$ and the approximate analytic solutions $\sum_{k=0}^m(\tilde{U}_{+,k}+\tilde{U}_{-,k})$ for $1\le m\le 3$. 
In Fig.~\ref{figure_phi_0} we compare the exact numerical solution $f$ with the approximate analytic solutions $\sum_{k=0}^m f_k$ for $0\le m\le 2$. 
Finally, Fig.~\ref{figure_r_error_phi_error} gives the accuracy for the wall radius $R$ in terms of the $\epsilon$-dependence of the relative errors $R^{-1}\sum_{k=0}^m R_k-1$ for $0\le m\le 3$ and the $\epsilon$-dependence of the integrated deviation of the exact numerical solution $f$ from the approximate analytic solutions $\sum_{k=0}^m f_k$ for $0\le m\le 2$.

\section{Discussion and Conclusions}\label{section_discussion}

We have proposed a new method of iterative approximate solutions for the spherical bounce which works for any continuously differentiable potentials in any number of dimensions. 
The zeroth-order and first-order approximations coincide with the thin-wall approximation of Coleman~\cite{Coleman:1977py} and all higher-order approximations are derived iteratively.

The iterative approximations have global features which distinguish them from more straightforward local approximations obtained via standard series expansions.
A local approximation typically works best near the center of the expansion, but its accuracy rapidly decreases far from the center.   
The situation is slightly better with matched series approximations, where several expansions centered at different points are glued at points between the centers by matching the first few derivatives of the solution. 
We gave an example of this matched series expansion in Sec.~\ref{subsection_power_series_approximation}, where we saw that its accuracy is not great for large asymmetry in the potential $U$.

On the other hand, having smaller errors
\begin{align}
  E_m=\frac{\int_0^\infty dr\,r^n\bigl(\sum_{k=0}^m f_k(r)-f(r)\bigr)^2}{\int_0^\infty dr\,r^n f(r)^2}, \label{f_error}
\end{align}
our iterative approximations better represent the exact solutions for a broad range of values of $r$, especially for $r>R$. 
We also note that even if the quantities $\sum_{k=0}^m f_k$ differ significantly from $f$ for small $r$, the presence of the Jacobian factor $r^n$ in the numerator in \eqref{f_error} makes these differences for $n\ge 1$ much less important than the corresponding differences for large $r$.
Compare Figs.~\ref{figure_phi_0} and \ref{figure_r_error_phi_error} in this regard.

Our method proceeds to higher orders iteratively with fast convergence and high accuracy.
Analysis of the problem from a new perspective demonstrates some universal properties of the bounce.
The method is not restricted to only certain types of potentials or dimensions of space.
For example, there is nothing special about the potential being a fourth-order polynomial or the space being three-dimensional for the successful application of the method to the example we investigated in Sec.~\ref{section_polynomial}.
We also note that the approximation works well beyond its intended range of applicability of small asymmetry of the potential. 
Compare Figs.~\ref{figure_u_phi}, \ref{figure_phi_0}, and \ref{figure_r_error_phi_error} in this regard.
The potential functions shown in the left part of Fig.~\ref{figure_u_phi} are precisely those for which the corresponding solutions and their approximations are shown in the right part of the Fig.~\ref{figure_u_phi} and in Fig.~\ref{figure_phi_0}.  
Although hardly any of these potential functions can be considered as having small asymmetry, the approximations in Fig.~\ref{figure_phi_0} are quite accurate. 

Once approximations for the classical bounce solution are known in the analytic form, the next obvious step is to compute the decay rate of the false vacuum.
The rate is the product of the exponential term given by the classical action of the bounce and the pre-exponential factor expressed in terms of functional determinants.
With the iterative method for spherical bounces developed in this paper, deriving corresponding approximations for the pre-exponential factor should be a relatively straightforward procedure.
Another promising direction is to develop a similar approximation method for the gravitational bounce; it would be interesting to see how the required modifications agree, in particular, with the findings of  Refs.~\cite{Coleman:1980aw}, \cite{Samuel:1991dy} and \cite{Masoumi:2016pqb}.

\section*{Acknowledgments}

I am grateful to Thomas W. Kephart for his participation in the early stages of this project and for useful discussions.

\end{document}